\documentclass[twocolumn]{emulateapj}

\shorttitle{Variable stars in IZw18.}
\shortauthors{Fiorentino et al.}

\begin{document}

   \title{Multi-Epoch HST Observations of IZw18: Characterization of
       Variable Stars at Ultra-Low Metallicities.\footnote{Based on observations with the
NASA/ESA Hubble Space Telescope, obtained at the Space Telescope Science
Institute, which is operated by AURA Inc., under NASA contract NAS 5-26555.}}

\author{G. Fiorentino\altaffilmark{1,2}, R. Contreras
  Ramos\altaffilmark{1}, G. Clementini\altaffilmark{1},
  M. Marconi\altaffilmark{3}, I. Musella\altaffilmark{3},
  A. Aloisi\altaffilmark{4}, F. Annibali\altaffilmark{5},
  A. Saha\altaffilmark{6}, M. Tosi\altaffilmark{1} \& R. P. van der
  Marel\altaffilmark{4}}

\affil{$^1$INAF-Osservatorio Astronomico di  Bologna, via Ranzani 1, 40127, Bologna, Italy} 
\affil{$^2$Kapteyn Astronomical Institute, University of Groningen, Postbus 800, 9700 AV Groningen, The Netherlands}
\affil{$^3$INAF-Osservatorio Astronomico di Capodimonte, via Moiariello 16, 80131, Napoli, Italy}
\affil{$^4$Space Telescope Science Institute, 3700 San Martin Drive, Baltimore, MD21218}
\affil{$^5$INAF-Osservatorio Astronomico di Padova, Vicolo dell’Osservatorio 5, I-35122 Padova, Italy}
\affil{$^6$National Optical Astronomy Observatory, P.O. Box 26732, Tucson, AZ 85726}

\begin{abstract}
  Variable stars have been identified for the first time in the very
  metal-poor Blue Compact dwarf galaxy IZw18, using deep multi-band
  (F606W, F814W) time-series photometry obtained with the Advanced
  Camera for Surveys (ACS) on board the Hubble Space Telescope (HST).
  We detected 34 candidate variable stars in the galaxy.  We classify
  three of them as Classical Cepheids, with periods of 8.71, 125.0 and
  130.3 days, respectively, and other two as long period variables
  with periodicities longer than a hundred days.  These are the lowest
  metallicity Classical Cepheids known so far, thus providing the
  opportunity to explore and fit models of stellar pulsation for
  Classical Cepheids at previously inaccessible metallicities.  The
  period distribution of the confirmed Cepheids is markedly different
  from what is seen in other nearby galaxies, which is likely related
  to the star bursting nature of IZw18. The long period Cepheids we have detected in IZw18 
  seem to indicate that massive stars at the metallicity of IZw18 (Z=0.0004) may cross the 
  instability strip long enough to be observed.

  By applying to the 8.71 days Cepheid theoretical Wesenheit ($V,I$)
  relations based on new pulsation models of Classical Cepheids
  specifically computed for the extremely low metallicity of this
  galaxy (Z=0.0004, Y=0.24), we estimate the distance modulus of IZw18
  to be $\mu_0$= 31.4 $\pm$ 0.3 (D=19.0$^{+2.8}_{-2.5}$ Mpc) for
  canonical models of Classical Cepheids, and of 31.2 $\pm$ 0.3 mag
  (D=$17.4^{+2.6}_{-2.2}$ Mpc) using over luminous models. The
  theoretical modeling of the star's light curves provides $\mu_0=31.4
  \pm$ 0.2 mag, D=19.0$^{+1.8}_{-1.7}$ Mpc, in good agreement with the
  results from the theoretical Wesenheit relations.  These {\it
    pulsation} distances bracket the distance of 18.2 $\pm 1.5$ Mpc
  inferred by \citet{aloisi07} using the galaxy's Red Giant Branch
  Tip.
\end{abstract}


\section{Introduction}
Discovered by \citet{searle73} more than thirty
years ago the Blue Compact Dwarf galaxy IZw18 remains one of the most
metal poor \citep[1/50 Z$_{\odot}$,][]{skillman93} and
intriguing objects in the local Universe.  The galaxy is very rich in
gas content and blue stars.  These two features both suggested that
IZw18 might be undergoing its first episode of star formation, thus
representing one of the best local analogs to primordial galaxies in
the distant Universe.

Several HST studies by different authors have tried to characterize
the evolutionary status of IZw18, with rather discordant
results. \citet{aloisi99} were the first to detect
Asymptotic Giant Branch (AGB) stars in the galaxy, and thus to
demonstrate that IZw18 is as old as several hundreds Myr, at least,
using HST Wide Field Planetary Camera 2 (WFPC2) data. \citet{izotov04} failed to detect red giant branch (RGB) stars in IZw18, based
on HST/ACS observations of the galaxy. However, \citet{momany05}
and \citet{tosi06}, based on a re-analysis of the same ACS
data set, suggested that IZw18 is older than previously believed, since
they found some evidence for the galaxy's RGB Tip around $I \sim $ 27
mag.  Our new time-series HST/ACS data (HST program Id. 10586, PI:
Aloisi) have allowed us to definitively solve the mystery on the age
of IZw18, by pinning down the galaxy's distance, using Classical Cepheids that we detected for the first time
in this galaxy. The distance to IZw18 inferred from the Classical
Cepheids is in very good agreement with the independent estimate
obtained from the galaxy's red giant branch (RGB) Tip ($D = 18.2 \pm
1.5$ Mpc), which is identified by \citet{aloisi07} at $I_0 =
27.27 \pm 0.14$ mag (for an assumed Galactic foreground extinction of
$E(B-V)$=0.032 mag, and shows that IZw18 is further away than
previously thought.  In this paper we present in detail the variable
stars we have identified in IZw18 and discuss in particular the
galaxy's Classical Cepheids.  Observations and data reduction are
described in Section 2. The techniques applied to identify candidate
variable stars and derive their periodicities are described in Section
3.  The properties of the IZw18 confirmed variable stars are presented
in Section 4, while the distance inferred from the Classical Cepheids
is presented in Section 5. Finally, our results and conclusions are
summarized in Section 6.

\section{Observations and data reductions}

Time series imaging of IZw18 in the F606W (broad $V$) and F814W ($I$)
filters was obtained with the Wide Field Channel of the Advanced Camera
for Surveys (ACS/WFC) on board the HST in 13 different epochs properly
spread over a time interval of 96 days from 2005 October to 2006 
January.  The time window and the scheduling of the single epoch
observations were specifically devised to allow the discovery and
characterization of Classical Cepheids with periods shorter than 100
days, if present.

Each epoch consisted of four optimally dithered single exposures per
filter, that we corrected for geometric distortion and co-added by
using the MultiDrizzle software, to produce single images per epoch
per filter, totaling about 35-37 minute integration time per epoch,
and total integration times, over the 13 epochs, of $\sim 7.7$ and
$\sim 7.3$ hours in F606W and F814W, respectively.  The proprietary
data were complemented by archival F555W ($V$) and F814W ACS/WFC images
of IZw18 from HST program Id. 9400 (PI Thuan), obtained in 2003,
May-June, over a time span of 11 days, thus gaining three and five
additional epochs in the F814W and F555W filters, respectively.  The
additional data allowed us to increase the number of available
phase-points on the light curves to 15 in F814W, 13 in F606W and five
in F555W and to extend the time interval beyond the 96 days spanned by
the proprietary observations. This is valuable to pin down periods
longer than the time window covered by the proprietary data, and to
reveal long period variability.  The log of the observations for both
proprietary and archival data is provided in Table\,1.

The severe crowding of the IZw18 images, the demand of achieving the
maximum possible depth in the color magnitude diagram (CMD), and the
need of processing the single epoch data to generate light curves for
the candidate variables, made the photometric reduction of our data
very challenging and time-consuming, and the use of PSF-fitting
photometry mandatory.  To produce the galaxy CMD, we built four
coadded images by stacking all single epoch images per filter for both
archival and proprietary data, separately. Then we performed PSF
photometry of each of these four coadded images using the
DAOPHOT/ALLSTAR package \citep{stetson87}.  The resulting CMD was
published by \citet{aloisi07}. It contains $\sim$ 2100 sources
having detection in all four deep images ($V$ and $I$ for both datasets),  
and shows the galaxy's RGB Tip at $I_{0}$=27.27 $\pm$ 0.14 mag.

\begin{figure}
\includegraphics[width=10cm]{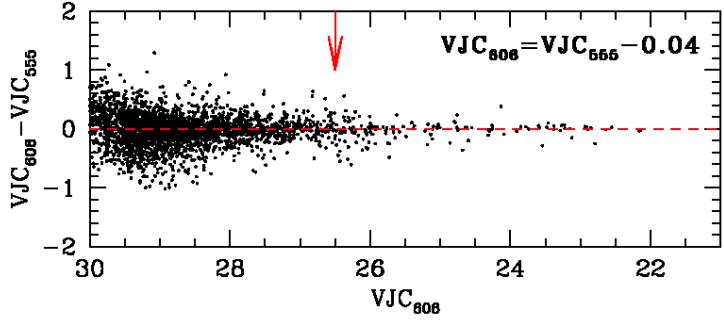}
\caption{Comparison between $V$ magnitudes obtained from the calibration
to the Johnson-Cousins system of the F606W and F555W magnitudes, separately. A zero point shift exists 
between  
the two $V$-calibrated photometries for magnitudes brighter than $V$=26.5 mag (arrow in the figure),
with the F606W $V$-calibrated magnitudes being on average 0.04 mag
brighter that the F555W ones.}
\label{zero}
\end{figure}

To produce light curves for the candidate variable objects we followed
the approach described in the $H_0$ cookbook by Turner (1997) and used
the MONTAGE package within DAOPHOTII/ALLSTAR/ALLFRAME \citep[][]{stetson87,stetson94} to build a master frame out of the four coadded images obtained
by \citet{aloisi07}. Then we performed DAOPHOTII/ALLSTAR/ALLFRAME
PSF photometry on each single epoch (15 $I$ and 18 $V$ images) using
as a reference the catalogue of sources obtained from the master
frame. The new photometry was calibrated to the F555W, F606W and F814W
HST Vegamag system by using the IZw18 photometric catalogue presented
in \citet{aloisi07} as a list of standard reference stars.
Finally, the calibration to the Johnson-Cousins photometric system was
obtained using the transformations given by \citet{sirianni05}
\citep[see][, for details]{aloisi07}. The F606W and F555W photometry
were calibrated to the Johnson-Cousins photometric system separately,
and the resulting standard $V$ magnitudes compared. The result of this
comparison is shown in Figure ~\ref{zero}, for $V-I$ colors redder
than 0.4 mag, which roughly corresponds to the average color inside
the classical instability strip (IS) in IZw18.  

A zero point shift was found to exist between the two $V$-calibrated
photometry for magnitudes brighter than V=26.5 mag.  (see arrow in
Figure ~\ref{zero}). In fact, the $V$ magnitudes derived from the
calibration of the F606W data ($V_{\rm F606W}$) are on average $0.04$
mag brighter than the $V$ magnitudes derived from the calibration of
the F555W data ($V_{\rm F555W}$). Since the F555W filter more closely
resembles the photometric passband of the Johnson-Cousins $V$
photometry, and the F606W fluxes might be somewhat overestimated
because of non-negligible gas contamination, the $V_{\rm F606W}$
magnitudes were corrected for this zero point and tied to the $V_{\rm
  F555W}$ ones. The photometric reductions with
DAOPHOTII/ALLSTAR/ALLFRAME produced a new $I, V-I$ CMD of IZw18 which
contains about 7000 sources, all having photometric errors
$\sigma_{V,I} < 0.5$ mag and $SHARP$ parameter\footnote{The SHARP
  parameter is related to the intrinsic angular size of the object
  image and measures the regularity and symmetry of the PSF stellar
  profile.} in the range from $-1$ to $+1$, and reaches $V\sim$ 30.0
mag, i.e. about half a magnitude fainter than the \citet{aloisi07}
CMD. The new CMD is shown in Figure ~\ref{cmd}.  We note that a
fraction of the additional stars in our CMD populate the blue plume of
IZw18, and that some of them appear to be bluer than the Padua tracks.
An eye inspection of a map of IZw18 confirms that these are likely gas
contaminated stars, usually located in gas-rich regions and/or gas
filaments.
 
\section{Variable star identification and period search}
We used four independent approaches to identify variable stars in
IZw18 and to select bona fide variables from false detections.  First
we applied to our proprietary F606W and F814W data sets the Optical
Image Subtraction Technique as performed within the package ISIS 2.2
\citep{alard00}.  This method has shown to be very powerful to detect
variables in highly crowded stellar fields such as the main body of
IZw18 \citep[see e.g.,][]{baldacci05}. With this approach we obtained a list of candidate variables and
differential flux light curves, on which to perform the period
search. However, the under-sampling of the ACS/WFC images revealed a
very challenging issue for ISIS. The number of sources flagged as
candidate variables was extraordinary large, making the check of the
individual objects extremely time-consuming and difficult. 

Moreover, inspection of the images revealed that ISIS had very often
failed to distinguish real stars from spurious sources, galaxies and
background noise.

We then used the new DAOPHOTII/ALLSTAR/ALLFRAME photometry of the
single-epoch data and an output parameter of ALLFRAME, the
Welch-Stetson variability index \citep[see ][ and the
$H_0$ cookbook]{welch93} to flag candidate variable stars in the catalogue of
measured sources returned by ALLFRAME. This second approach allowed us
to pick up 66 candidate variables. All of them were already present in
the list of variables detected by ISIS. However, by using the light
curves in magnitude scales produced by ALLFRAME, it was much easier
and straightforward to check the candidate variables than by working
with the differential flux light curves of ISIS, where we lacked
information on the actual amplitude of the light variation.

As a third approach, PSF-fitting photometry of the single epoch data
was performed with DoPHOT \citep[][]{schechter93} and we used
the package procedures, as discussed in \citet{saha90}, to
search for periodic variables.  Briefly, once we had star magnitudes
at several epochs, variable star candidates were detected as the
sources whose data points had a large scatter in the $\Delta$ MAG vs
MAG plane (where $\Delta$ MAG is the difference between magnitudes at
two different epochs) and good parameters of sharpness and
goodness-fit criteria ($\chi ^2$). To make the flag as candidate
variable more robust, the empirical scatter was compared with the
average error estimated by DoPHOT. This procedure was applied to
several frame pairs. The scattering data points were then checked by
eye.  They often revealed spurious variable sources, due to the
difficulty of performing a reliable PSF-fitting photometry in IZw18.

Finally, identification of candidate sources was also performed by
visual inspection of the $\chi ^2$ F606W and F814W images by one of us
(A.S.).  In this approach each pixel is the $\chi ^2$ value calculated
from the corresponding pixels in the co-registered target images at
all epochs.  The stability of the HST/ACS PSF, combined with the
careful registration of the images makes it possible to visualize the
variable objects very effectively.  The co-registered images in the
same passband can be stacked as a data cube, with each layer in the
``z" direction as one of the time epochs. An estimate of the noise in
each pixel of this data cube is obtained from the Poisson statistics
and the read noise of the detectors.  From the variance of each pixel
along the ``z" direction, and the noise estimates as above, a $\chi
^2$ is obtained, thus producing a $\chi ^2$ image. All of the definite
variables identified in this paper are clearly visible as pixel
groupings with high values in the $\chi ^2$ images. We have not
attempted a detailed PSF analysis of these $\chi ^2$ ``stars", but
visually, the majority of these enhancements appear as compact and
``round" structures, with sizes consistent with that of single stars.
Note that the level of enhancement corresponds to the statistical
significance for variability, and objects that are relatively very
faint can show up prominently. Note also that several $\chi ^2$
enhancements which are consistent with the ``footprint" of single
stars, appear in the central body of IZw18. In these areas, the
crowding makes it impossible to perform any photometry, or even
de-blend individual stars in the actual images. It is thus unfortunate
that even though the $\chi ^2$ maps show the presence of variable
stars so cleanly in such confused regions, they remain inaccessible to
further analysis.  The $\chi ^2$ image approach is an extension of the
``difference image" method for identifying variables. It would
eliminate the need to compare images pair-wise, allowing rather to
look for statistically significant variability in multi-epoch data,
while preserving the information provided by the correlation between
neighboring pixels due to the PSF. Clearly, the method bears further
investigation and development.

\begin{figure}
\includegraphics[width=10cm]{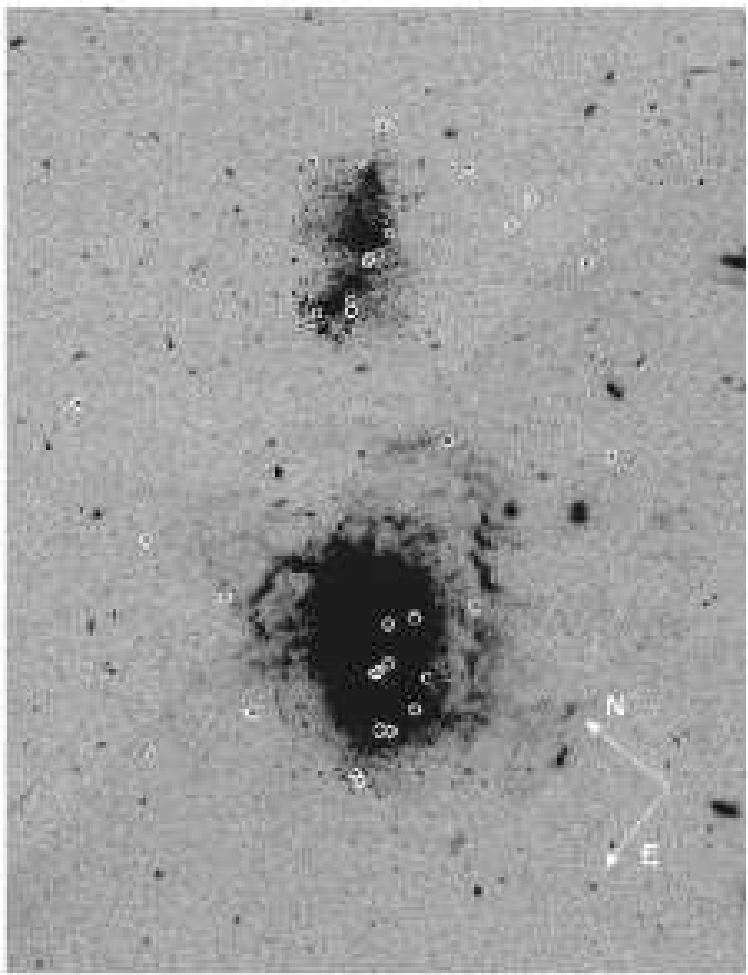}
\caption{Image of IZw18 showing the location of all 34 candidate
  variables (open circles), including the five confirmed variable
  stars.}
\label{map-cand}
\end{figure}

Candidate variable stars found by the four independent procedures
described above were counter identified by coordinates, returning a
final catalogue of 34 bona fide candidates showing light variations
larger than three times the photometric error at their average
magnitude, that we retained for further analysis. Nineteen out of the
34 candidates lie on and around the main body of IZw18, the remaining
15 ones on the galaxy secondary body. Figure ~\ref{map-cand} shows the
location of the 34 candidate variables on a map of IZw18.

\begin{figure}
\includegraphics[width=10cm]{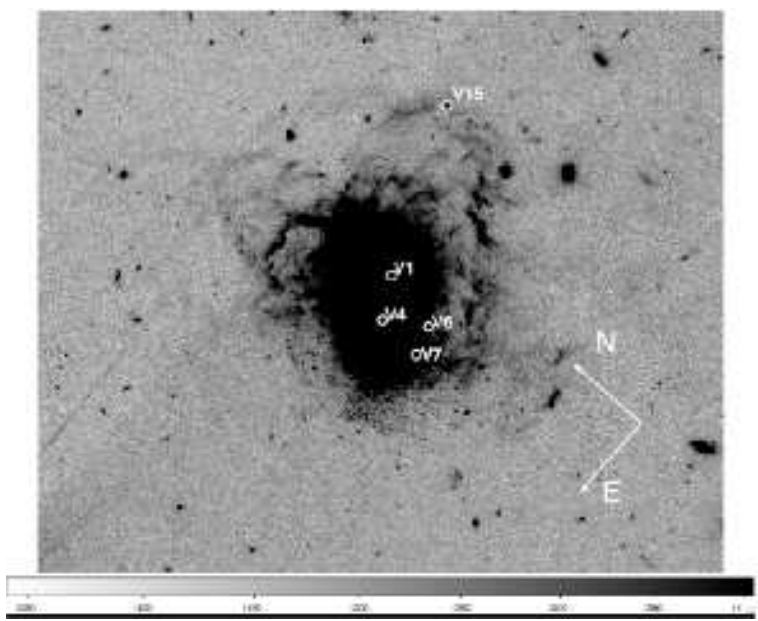}
\caption{Image of the main body of IZw18 showing only the location
  of the five confirmed variable stars (open circles).}
\label{map-cep}
\end{figure}

Period search and study of the light curves for the bona fide
candidate variables were performed using an iterative procedure that
we applied to the F814W, F606W and F555W instrumental magnitudes,
separately.  A first guess of the star's periodicity was obtained
using the Phase Dispersion Minimization (PDM) algorithm, within the
IRAF environment, making a quick search over a large time interval
ranging from 2-3 days to about 200 days. The period refinement was
then obtained with the Graphical Analyzer of Time Series (GRaTiS), a
private software developed at the Bologna Observatory \citep[see e.g.,
][]{clementini00}, which uses both the Lomb periodogram
\citep{lomb76,scargle82} and the best fit of the data with a
truncated Fourier series \citep[e.g., ][]{barning63}.  First we run GRaTis on the
15 epochs of the 2003+2007 F814W data set deriving a period that we
used to phase the F555W and F606W light curves, and iteratively
improved by comparing the light curves in the three different
pass-bands. Candidate variables showing the same consistent
periodicity in all the three bands were confirmed as bona fide
variable stars, while the other stars remained potential candidates.
This procedure allowed us to select, among the 34 candidates, 5
variable stars with consistent light curves in all the three
filters. These five bona fide variables are listed in Table\,2 and
plotted as red filled triangles in Figure ~\ref{cmd}.  Their position
on the body of IZw18 is shown in Figure ~\ref{map-cep}. Four of them
(V1,V4, V6 and V7) lie on the galaxy main body.  The fifth one, V15,
is on a gas shell surrounding the main body and pointing towards the
direction of IZw18 secondary body. The 5 confirmed variables are
further discussed in Section 4.

For the remaining 29 objects (see Table\,3) we only have an
identification, the mean $V$, $I$ magnitudes provided by ALLFRAME, and
a rough indication of the light variation in the F606W and F814W
bands, since several of these objects lie in areas where crowding
makes it impossible to perform a reliable photometry of the single
epoch data.  We also note that although many of the objects in
Table\,3 show significant multi band light variations, for most of them
the $V$ light curve does not match the $I$ curve for the same
period. This is likely due to crowding that makes the
counter identification of an object in different photometric bands
rather uncertain, and to stellar blending, which affects in different
ways the star magnitude, depending on the photometric band. It could
also be possible that some of these candidate variables are in
unresolved stellar associations. The candidate variable stars are
shown as orange filled circles in Figure ~\ref{cmd}.

\section{IZw18 confirmed variable stars}

The time series data of the five bona fide variable stars were
calibrated to the Johnson-Cousins photometric system following the
general procedure described in Section 2 and properly taking into
account the color variation of the variable stars during the pulsation
cycle. In practice, we associated to each phased F555W and F606W data
point the F814W value read, at the same phase, from the model best
fitting the F814W data, and to each phased F814W data point the F606W
value at same phase read from the model best fitting the F606W
data. The $V$ magnitudes derived from the F606W photometry were
corrected for the 0.04 mag zero point shift before combining them to
the F555W calibrated photometry to derive the average magnitudes of
the variable stars over the pulsation cycle. The time-series $V,I$
photometry of the 5 confirmed variables is provided in Table\,4 which
is published in its entirety in the electronic edition of the journal.

\begin{figure}
\includegraphics[width=10cm]{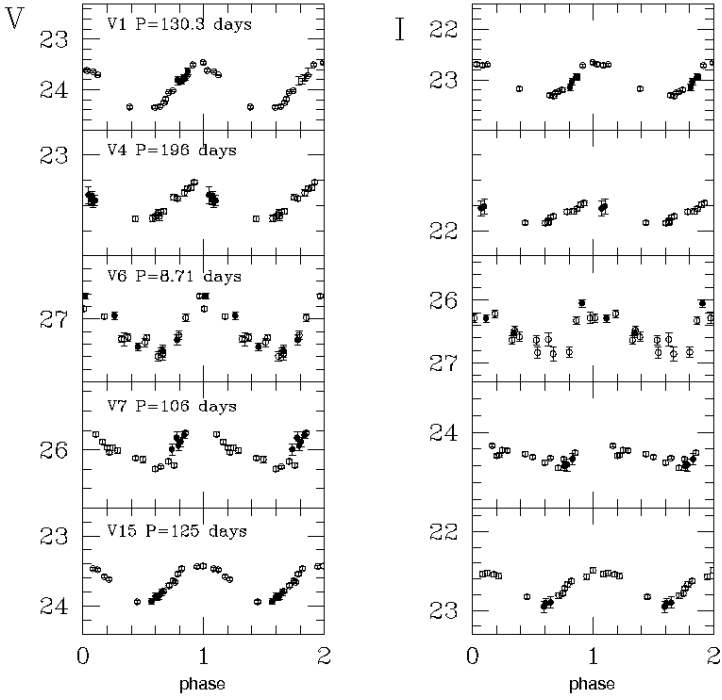}
\caption{$V$ (left) and $I$ (right) light curves of the confirmed
  variable stars in IZw18. Stars are labeled according to
  Table\,2. The star's periods are indicated. Open and filled circles correspond
  to our 2007 proprietary data, and to the archival data, respectively.}
\label{lc}
\end{figure}

Results from the analysis of the 5 confirmed variables are summarized
in Table\,2, where for each star we provide identification number,
coordinates, period, time of maximum light of the $V$ data,
intensity-averaged mean magnitudes and colors, amplitudes in the $V$
and $I$ bands of the Johnson-Cousins photometric system, and amplitude
ratios.  Figure ~\ref{lc} shows the Johnson-Cousins $V,I$ light curves
of the five confirmed variables, where data have been folded according
to the periods and times of maximum light listed in Table\,2.  Error
bars show the intrinsic errors of the PSF photometry, as computed by
ALLFRAME, on the single epoch measurements.  For the faintest variable
star (V6) they are of the order of 0.09 mag in $V$ and 0.12 mag in
$I$.  We note that only for V6 the pulsation cycle is repetitively
covered by our observations, thus making the period determination
robust. We estimate an uncertainty of the order of $\pm$ 0.08 days on
the period of this star. The other 4 confirmed variables all have long
periodicities exceeding the time window covered by the proprietary
data. Although the addition of the archival data allowed us to improve
the period determination, for these stars we find more uncertain
periods than for V6.
 
For the longer period variables we first used the Lomb algorithm
within GRaTiS to explore a period interval ranging from a few days up
to 300-500 days, and chose, as usual, the most likely period as that
corresponding to the highest peak in the star's power spectrum. The
period definition was then refined using the truncated Fourier
algorithm.  In general, given the smooth and sinusoidal shape of the
light curves, 2 harmonics were sufficient to best fit the data.
According to the width of the highest peaks in the power spectra of V1
and V15, the accuracy of the derived periodicities is of the order of
$\pm$ 1day.  However, it should be noted that both these two stars
also have lower peaks in the power spectra corresponding to P=113 and
152 days for V1, and P=110 and 146 days for V15. We are inclined to
attribute these alternative periodicities to the windowing of the
data. However, only further time-series covering much longer time
windows could allow us to constrain the actual periods.

The period definition is more uncertain for V4 and V7.  Indeed, the
Lomb periodogram would favor a period around 165 days for V7, while
the truncated Fourier series favors a shorter period of 106 days.  Our
data do not allow us to distinguish between these two periodicities.
For practical purposes we adopted the shorter period to fold the light
curve data, but this period should be taken with caution.

Similarly, the very long period of V4 (196 days) significantly exceeds
the time window covered by the proprietary data (96 days), and after
addition of the archival data yet we apparently covered only the
rising branch of the star light curve, thus rendering the period
determination very uncertain. The highest peak in the power spectrum
of V4 is double peaked, with two equally possible periodicities of 187
and 196 days. Moreover, several other longer periods exceeding the 200
days are also possible. We used the 196 period to fold the ligh curve
data; however, as with V7, the period of V4 should be considered only
a tentative estimate.

In order to classify the bona fide variable stars in types we have
used the following three diagnostics: (i) the star's pulsation period;
(ii) the shape, the morphology, and the amplitude ratios of the light
curves; (iii) the location on the galaxy's CMD in comparison with
evolutionary tracks by the Padua group \citep[][; hereafter Padua94]{fagotto94} for the mass range from 5 to 60 M$_{\odot}$ (solid curves in
Figure ~\ref{cmd}). The Cepheid candidates were also compared with the
theoretical boundaries of the classical instability strip (IS)
specifically computed for the extremely low metallicity of IZw18
(Z=0.0004, Y=0.24, Marconi et al. 2010 accepted on ApJ, hereafter M10), and for two
different choices of the mixing length parameter, namely, $\alpha$=1.5
and 2 (solid and dashed lines in Figure ~\ref{cmd}).

Three out of the five variable stars, namely V1, V6 and V15, have
light curves closely matching those of Classical Cepheids.  Figure
~\ref{cmd} shows the location of the five variable stars on the $I,
(V-I)$ CMD of IZw18 obtained by the present analysis.  The confirmed
variable stars (filled triangles in Figure ~\ref{cmd}) are plotted
according to their intensity-averaged mean magnitudes and colors,
obtained by weighting over the star pulsation cycle and, with the
exception of V4 and V7, occupy a region of the galaxy CMD well
confined in color.  Both V1 and V15 in the CMD lie well inside the
boundaries of the theoretical IS for Cepheids at the metallicity and
reddening of IZw18, while the faintest of the three variables, star
V6, lies close to the blue edge of the predicted IS, but still within
the 1$\sigma$ confidence level.  We note that the $V$ amplitude of V6
seems to be slightly small compared to the $I$ amplitude (see columns
8, 10 and 11 of Table\, 2). The blue color and the slightly small $V$
amplitude could be explained if the star were contaminated by an
unresolved blue companion or by gas.
$\chi ^2$ and sharpness parameters of V6 do not seem to differ from those of a regular single
stellar source, but the star is clearly located in a gas rich region (see
Figure ~\ref{map-cep}), thus could easily be contaminated by gas. The
large throughput of the F606W filter towards emitting gas wavelengths
along with V6 intrinsic faintness could thus have conspired to render
the star $V-I$ color bluer than normal.

In summary, the different diagnostics all consistently suggest a
classification as Classical Cepheids for V1, V6 and V15. However, of
the three stars only V6 (P=8.71 days) has periodicity similar to the
typical period ranges commonly observed for Classical Cepheids in and
outside the Milky Way. This star also safely falls below the period of
P$\sim$ 10 days, where a break has been suggested to occur in the
Period $-$ Luminosity relation of Classical Cepheids 
\citep[see, e.g.][]{ngeow04,ngeow09,sandage04}.
V1 and V15 instead have rather long periods, which, so far, have been
observed only very rarely among Classical Cepheids.  In a recent
paper, \citet{bird09} list the very long period Classical Cepheids
known so far, which they name ultra-long period (ULP) Cepheids.  \citet{bird09} list includes a total number of 18 ULP Cepheids with periods
in the range from 83.0 to 210.4 days, distributed over five galaxies,
namely, the Large and Small Magellanic Clouds, NGC300, NGC6822, and
IZw18, referring to preliminary results for V1 and V15 that we
published \citep[][]{fiorentino08}. Inspection of the literature light
curves compiled by \citet{bird09} reveals that those of the ULP Cepheids
seem to be more sinusoidal and to have a stronger dependence on
wavelength than their classical ``short'' period counterparts.  In
most of the cases the $I$ band light curves show a sort of {\it
  plateau} rather than the peak at maximum light observed in the
$V$-band light curves. However, the amplitude ratios, A$_I$/A$_V$, do
not seem to differ significantly from those of Classical Cepheids in
standard period ranges (see last column of Table~2).

As already noted in \citet{bird09},  the ``theoretical'' interpretation of 
ULP Cepheids like the bright and long period
Cepheids detected in IZw18 poses some problems. In fact the current
set of nonlinear convective pulsation models fails to predict
pulsation periods longer than about 90 days at this metal content for
masses in the typical range of Classical Cepheids ($3 \leq $M$ \leq
13$ M$_{\odot}$; see discussion in M10) and current evolutionary calculations
do not predict the existence of a blue loop at so high luminosity
levels (see discussion below). Nevertheless, we emphasize the
importance of studying Classical Cepheids at the very low metallicity
of IZw18 to constrain the still unexplored properties of the
period-luminosity (PL) relation in very poor metallicity regimes. This
topic is addressed in M10, where we have combined the new pulsation
models computed to study the IZw18 Cepheids with previous sets of
models at higher metal abundance \citep[][, and references therein]{mmf05} to constrain the metallicity dependence of the Cepheid PL
relation in the low metallicity regime (Z from 0.0004 to 0.008).

The other two confirmed variables in IZw18, V4 and V7, have both
rather red colors. Indeed, V7 is the reddest of our confirmed
variables with $\langle V-I \rangle$ =1.71 mag. The star lies about 3
magnitudes above the RGB Tip of IZw18 in the expected region of TP-AGB
stars. The position in the CMD, the long period and the shape of the
light curve suggest that the star could be a long period variable
(LPV) of Mira type.

The fifth variable, star V4, is one of the brightest and reddest
objects in the IZw18 CMD. Only a portion of the star light curve is
covered by our observations, thus both period and amplitudes of the
light variation are very uncertain.  Although the brightness and shape
of the light curve could suggest the classification of V4 as a
Cepheid, the very red color places the star well outside the typical
color range of Classical Cepheids. The uncertainty of the period,
which anyway appears to be extraordinarily long for a Classical
Cepheid, and the rather red color, which seems intrinsic and not due
to contamination by a red companion, suggest to adopt a conservative
classification as LPV, and to not use the star to estimate the
distance to IZw18.

The theoretical evolutionary tracks shown in Figure ~\ref{cmd} span
from the post-Main Sequence up to the early Asymptotic Giant Branch
evolutionary phase, thus covering both the phase of central Helium and
shell Hydrogen burnings. In this phase an intermediate-mass star
crosses the pulsation IS while performing the blueward excursion
usually called ``blue loop'', and thus can be observed as Classical
Cepheid.  According to the Padua94 models, the evolutionary track that
within the uncertainties most closely reproduces the range in
magnitude and color spanned by the Cepheid with P=8.71 days ($26.2 <I<
26.8$ mag, and $0.5 <V-I<0.7$ mag, respectively) corresponds to $M=6
M_{\odot}$ (red line in Figure ~\ref{cmd}). The Padua tracks include a
mild overshooting resulting in {\em blue loops} brighter by about 0.25
dex than tracks withouth overshooting such as the ones predicted by
canonical evolutionary scenarios [see also
http://astro.df.unipi.it/SAA/PEL/Z0.html]. In the canonical
assumption, the same luminosity level is attained for higher stellar
masses closer to $M=6.5 M_{\odot}$.

On the other hand, the brighter Cepheids should have masses larger than 15 M$_{\odot}$.  However,  the 20 M$_{\odot}$ track crosses the 
IS in the very fast phase of central Helium exhaustion, and the coarse distribution in mass of models between 15 and 20 M$_{\odot}$ does not allow
us to distinguish the largest mass that performs the  ``blue loop''
within the IS. The long period Cepheids we have detected in IZw18 
  seem to indicate that massive stars at the metallicity of IZw18 (Z=0.0004) may cross the 
  instability strip long enough to be observed. Whether this occurs during the ``blue loop" or not should be 
  further investigated from a theoretical point of view by both stellar evolution and pulsation models.  
  Blue loops are not predicted for massive stars by current stellar models, 
  but are invoked by \citet{bird09} for  the case of  HV829,  proposed to be a ``second crossing" ULP Cepheid in the SMC (Z=0.004).

\section{The distance to IZw18 from Classical Cepheids}

In the following we apply theoretical Wesenheit (WPL) relations, based
on the new pulsation models computed by M10, to the Classical Cepheid
with $P=8.71$ d (V6), to constrain the distance modulus of IZw18.  
We do not apply the same relation to the Cepheids with longer periods because pulsation models fail to predict
such very long periods at this chemical composition. However,
extrapolation of the predicted WPL relations to the very long periods
of V1 and V15 seems to well reproduce the data (see M10 for details).
Moreover, given the small number of Cepheids identified in IZw18, it
is not safe to use the PL relations, which are known to provide
reliable results only when applied to statistically significant
samples, and we prefer to use instead pulsation relations that can be
applied to individual Cepheids, such as the PLC and the WPL \citep{madore82} relations. In particular, the latter has the advantage of being
reddening-free by definition \citep[see][, for
details]{madore82,caputo00}.

A further method to measure the distance modulus of IZw18, based on
our new pulsation models, is the light curve model fitting technique
\citep[see e.g.,][and reference therein]{bono02,marconi05,keller06,natale08}.  In the following subsections we will discuss and use these
two methods, separately, and will then compare their results with the
distance to IZw18 derived by \citet{aloisi07} from the RGB Tip
method.
\subsection{Theoretical Wesenheit relations}

Updated reddening free theoretical ($I, V-I$) WPL relations have been
derived from our new theoretical pulsation models with Z=0.0004 and
Y=0.24 (M10) using both a canonical and a non-canonical (overluminous
by about 0.25 dex in $\log L/L_\odot$) assumption on the evolutionary
Mass-Luminosity relation (see M10 for details):\par
$<M_I>-$1.54($V-I$)=$-$2.58($\pm$0.12)$-$3.49($\pm$0.03)log$P$+0.84($\pm$0.10)$\log{L/L_C}$\par\noindent
with $\sigma$=0.12 mag and $\log{L/L_C}$=0.25 dex, if we use
overlouminous models, thus including both canonical and overluminous
solutions.
  
By applying this new WPL relation to V6 we obtain $\mu_0$=31.4$\pm$0.3
mag (D=19.0$^{+2.8}_{-2.5}$ Mpc) in the canonical assumption, and
$\mu_0$=31.2 $\pm$0.3 mag (D=$17.4^{+2.6}_{-2.2}$ Mpc) from
overluminous models.  The distance moduli obtained by the WPL
relations bracket the RBG Tip modulus of IZw18 obtained by \citet{aloisi07} ($\mu_0$=31.30$\pm$0.17 mag), and all three values are
consistent within the errors.

\subsection{Light curve model fitting}

The nonlinearity of our pulsation code allows us to predict the
variation of all the relevant physical quantities of a variable star
along the pulsation cycle and, in particular, to model the shape and
morphology of the light curves of individual Cepheids. This capability
can be used to theoretically fit the observed light curves, thus
providing another estimate of the stellar parameters and the
corresponding distance modulus of the host galaxy.  This technique was
already applied with success to both RR Lyrae stars and Classical
Cepheids \citep[see e.g.,][and reference therein]{bono02,marconi05,keller06,natale08}.  Here, we
have applied the method to V6, using the new pulsation models by M10.
In particular, we explored the mass range from 5.9 to 6.8 M$_{\odot}$,
as suggested by the comparison with different sets of canonical and
non-canonical evolutionary models, and computed isoperiodic sequences
of pulsation models by varying both the luminosity and the effective
temperature.

\begin{figure}
\includegraphics[width=10cm]{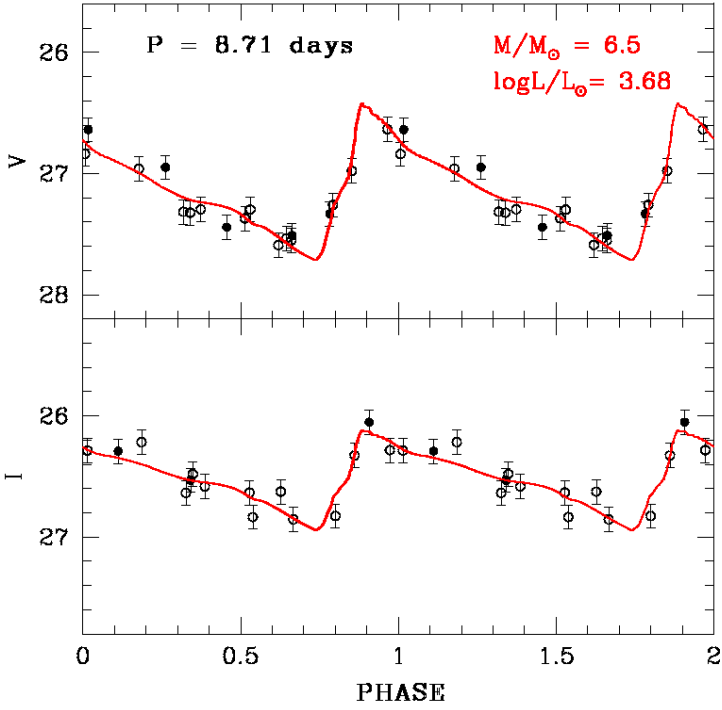}\label{fit}
\caption{Theoretical model fitting of the $V,I$ light curves of the P=8.71 days Cepheid (star V6).  The stellar parameters of the
  theoretical models (solid lines) best reproducing the observed light
  curves are labelled. Symbols are as in Figure 5.}
\end{figure}

The results of the fitting procedure are shown in Figure ~\ref{fit},
where we compare the observed light curves (filled circles) to our
best fit model (solid lines), that is the model minimizing the mean
residuals with respect to the observed curves in both $V$ and $I$
bands.  The best fit model has input parameters: M=6.5 M$_{\odot}$,
$\log{L/L_{\odot}}=$3.68 dex and T$_e=$5855 K, corresponding to
apparent distance moduli of 31.5 $\pm$ 0.2 mag in both
bands, where the uncertainty includes both the internal dispersion of
the fit and the observational errors.  
Therefore, by correcting the obtained $\mu_V$ and $\mu_I$ values 
for the assumed Galactic foreground extinction we find $\mu_0$=31.4 $\pm$ 0.2 mag.
The distance modulus derived with the model fitting technique is in
good agreement with the results obtained from the application of the
WPL relations and the RGB Tip method.

Taking into account all the pulsation models that fit the $V,I$ light
curves within a mean residual $\le 0.13$ mag, we can estimate the
theoretical uncertainties on the intrinsic stellar parameters of the
best fit model.  We find that luminosities from 3.64 to 3.69 dex in
$\log{L/L_{\odot}}$ and effective temperatures from 5800 to 6000 K are
allowed, for our selection of masses (5.9 to 6.8 $M_{\odot}$). These
masses, luminosities, and effective temperatures are in good
agreement, within the uncertainties, with the values predicted by
stellar evolution models (see discussion in Section 4). In particular,
our result that the best fit model corresponds to a stellar mass of
6.5 $M_{\odot}$ seems to support a canonical evolutionary scenario.

\section{Discussion and Conclusions}

We have identified and derived periods for five variable stars in the
blue compact dwarf galaxy IZw18, among a list of 34 candidates, based
on time series images of the galaxy obtained with the Advanced Camera
for Surveys on board the HST. Three of the variable stars are
Classical Cepheids, the remaining 2 are very red objects that we
tentatively classify as long period variables.

The distance to IZw18 was estimated from the Classical Cepheids using
theoretical $V,I$ Wesenheit relations specifically computed for the
very low metal abundance of the galaxy, and by theoretically fitting
the observed light curves of the 8.71 days Cepheid.  The Wesenheit
relations provide an average distance modulus for IZw18 of 31.4
$\pm$0.3 mag (D=19.0$^{+2.8}_{-2.5}$ Mpc) adopting a canonical ML
relation and of 31.2 $\pm$0.3 mag (D=$17.4^{+2.6}_{-2.2}$ Mpc) when a
overluminosity of 0.25 dex (in $\log{L_/L_C}$) is assumed for each
given mass.  The theoretical modeling of the 8.71 days Cepheid light
curves provides: 31.4 $\pm$0.2 mag (D=19.0$^{+1.8}_{-1.7}$ Mpc).

The Cepheid-based distance estimates are fully consistent within the
errors with the distance modulus of 31.30 $\pm$ 0.17 mag, obtained by
\citet{aloisi07} from a totally independent distance indicator,
namely, the RGB Tip of IZw18.
   
Classical Cepheids and the RGB Tip of IZw18 both consistently indicate
that the galaxy is much further away than previously thought, in turn
supporting the suggestion by \citet{aloisi07} that IZw18 is at
least as old as about 2 Gyrs.

\begin{figure}
\includegraphics[width=10cm]{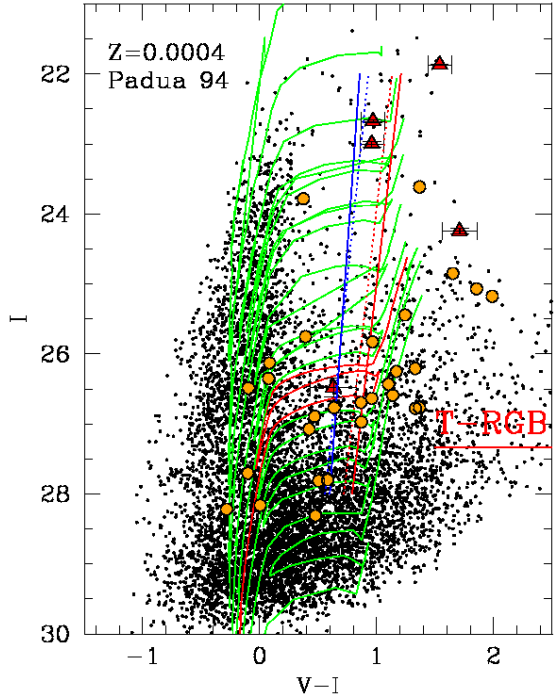}
\caption{$I, V-I$ CMD of IZw18 obtained in the present study.
  Points have not been corrected for Galactic extintion.  Filled red
  triangles show the five confirmed variable stars detected in IZw18,
  which are plotted according to their intensity-averaged mean
  magnitudes and colors. Filled orange circles show 29 candidate
  variables for which we could not derive reliable periods.  The solid
  green lines are the evolutionary tracks by \citet{fagotto94} for
  masses of 4, 5, 7, 9, 12, 15, 20, 30 and 60 M$_{\odot}$ and
  Z=0.0004. The evolutionary track that best fits the variable at
  $\langle I \rangle$= 26.48 mag and $\langle V-I \rangle$= 0.63 mag
  (star V6) is shown in red, and corresponds to a 6 M$_{\odot}$. Solid
  and dashed lines represent the theoretical boundaries of the
  Instability Strip (IS) computed for mixing length values of 1.5 and
  2, respectively (see Section 4 for further details). Both
  evolutionary tracks and boundaries of the theoretical IS were
  reddened for a Galactic foreground extinction of $E(B-V)$=0.032 mag
  (corresponding to A$_V$=0.106 and A$_I$=0.062 mag, respectively).
  The horizontal red line at $I$ = 27.33 mag shows the location of
  IZw18 RGB Tip (T-RGB), according to \citet{aloisi07} reddened
  value.}
\label{cmd}
\end{figure}

The number of Classical Cepheids we were able to identify in IZw18 is
undoubtedly very small, compared to the galaxy richness in gas and
blue stars. Also noticeable is the lack in IZw18 of Classical Cepheids
with periods intermediate between the 10 and 100 days. As a matter of
fact the CMD of IZw18 does not appear to containn many stars within
and/or close to the theoretical boundaries of the IS (see Figure
~\ref{cmd}). As an additional check we searched for variability all
the stars that lie in this region of the CMD, but did not succeed to
identify any further reliable candidate.

The gap in periods, as well as in luminosities, of the Cepheids in
IZw18 could be related, as discussed in \citet{aloisi07}, to the
starbursting nature of IZw18. In fact, although very active in very
recent epochs, the galaxy may lack a significant Star Formation (SF)
at the epochs when the missing Cepheids should have formed. Indeed,
once we scale the SF history derived by \citet{aloisi99} from
their assumed distance of 10 Mpc for IZw18 to the 18.2 Mpc distance
derived by \citet{aloisi07}, we expect few stars in the mass range
from 6 to 20 M$_{\odot}$ (corresponding to pulsation periods from 10
to 120 days) currently into the IS. A detailed derivation of the SF
history of IZw18 from the new data is in progress (Annibali et
al. 2010, in preparation).
 
\acknowledgments 

Financial support for this study was provided by COFIS ASI-INAF
I/016/07/0. Support was also provided by NASA through grants
associated with program GO-10586 from the Space Telescope Science
Institute (STScI), which is operated by the Association of
Universities for Research in Astronomy, Inc., under NASA contract
NAS5-26555. We thank Michele Cignoni for his help in homogeneously
transforming the stellar models to the observational plane, and Paolo
Montegriffo for making GRaTiS available and for continuous software
support.

\bibliographystyle{apj} 
\bibliography{variabili}

\begin{table}
\begin{center}
\caption[]{Log of the observations}
	\begin{tabular}{lcccc}
	   \hline
	    \hline
	   \noalign{\smallskip}
${\rm ~~~~~Dates}$ & ${\rm Texp}$ & ${\rm HJD^{a}}$          & ${\rm Filter}$& ${\rm Program~ID}$ \\ 
${\rm ~~~~~~UT}$   &${\rm sec}$  & ${\rm (-2450000)}$  &	        &     \\
	    \noalign{\smallskip}
	    \hline
	    \noalign{\smallskip}
	2003-05-26  &	  8100  &  2786.129397	                  &  F555W  & GO 9400\\  
	2003-05-29  &	  8100  &  2788.995991                    &  F814W  &  ''\\ 
	2003-05-30  &	 11120  &  2790.301071	                  &  F555W  &  ''\\  
	2003-05-31  &	  8100  &  2790.995864                    &  F814W  &  '' \\ 
	2003-06-01  &	  8100  &  2791.995813	                  &  F555W  &  '' \\  
	2003-06-03  &	  8100  &  2793.795690	                  &  F555W  &  '' \\  
	2003-06-05  &	  8100  &  2795.928907                    &  F814W  &  '' \\ 
	2003-06-06  &	  8100  &  2796.879372	                  &  F555W  &  ''\\  
	2005-10-02  &	  2140/2140  &  3646.074922/3646.200173   &  F606W/F814W &  GO 10586\\  
	2005-10-29  &	  2140  &  3673.354491	          &  F606W  &        '' \\  
	2005-11-04  &	  2140/2196  &  3679.215186/3679.282332  &  F606W/F814W  &   ''  \\  
	2005-11-08  &	  2140/2196  &  3683.344260/3683.411429  &  F606W/F814W  &   '' \\  
	2005-11-10  &	  2140/2196  &  3685.208977/3685.276123  &  F606W/F814W  &   '' \\  
	2005-11-13  &	  2140/2196  &  3688.11391273688.189299  &  F606W/F814W  &   '' \\  
	2005-11-18  &	  2140/2196  &  3692.577572/3692.653076  &  F606W/F814W  &   '' \\  
	2005-12-03  &	  2140/2196  &  3708.326761/3708.381510  &  F606W/F814W  &   '' \\  
	2005-12-10  &	  2140/2196  &  3714.535979/3714.649539  &  F606W/F814W  &   '' \\  
	2005-12-20  &	  2140/2196  &  3725.390374/3725.456858  &  F606W/F814W  &  '' \\ 
	2005-12-25  &	  2140/2196  &  3730.251196/3730.318190  &  F606W/F814W  &  '' \\ 
	2006-01-01  &	  2140/2196  &  3737.110567/3737.177722  &  F606W/F814W  &  '' \\ 
	2006-01-06  &	  2008/2056  &  3742.035879/3742.103508  &  F606W/F814W  &  '' \\ 
\hline
	 \end{tabular}
         \end{center}
\normalsize
$^{a}$ HJD is the Helicocentric Julian Day of observation at mid exposure.\\
	    \end{table}

\linespread{0.7}
\begin{table*} \caption{Identification and properties of the confirmed variable 
stars in IZw18.}
\begin{center}
\small
\begin{tabular}{lcccclccccc}
\hline
\hline
	   \noalign{\smallskip}
          &   ${\rm \alpha ^b}$ & ${\rm \delta ^b}$ &  ~~~${\rm P}$ & 
	    ~~~${\rm Epoch(max)}$  &  & & &
	    \\
            ${\rm Name^a}$& ${\rm (2000.0)}$& ${\rm (2000.0)}$&  ~${\rm (days)}$& ($-$2450000) &${\rm Type^c}$ &${\langle V\rangle}$ & 
	     ${\rm A_V^d}$ &${\langle I\rangle}$&  ${\rm A_I^d}$ & ${\rm A_I}$/${\rm A_V^d}$ \\
	   \noalign{\smallskip}
            \hline
            \noalign{\smallskip}
V1  &   9:34:1.94  & 55:14:25.87 & 130.3    &~~~~~~3726.00 &  CC   &23.96  &0.91  & 23.00   & 0.65 & 0.71\\
V4  &   9:34:2.30  & 55:14:24.17 & 196.0    &~~~~~~2777.00 &  LPV  &23.41  &$\ge$0.29  & 21.87   & $\ge$0.15 & 0.52\\
V6  &   9:34:2.07  & 55:14:21.23 & ~~~8.71  &~~~~~~3685.15 &  CC   &27.11  &0.85  & 26.48   & 0.63 & 0.74 \\
V7  &   9:34:2.33  & 55:14:20.49 & 106.0    &~~~~~~3662.00 &  LPV  &25.96  &$\ge$0.49  & 24.25   & $\ge$0.25 & 0.54 \\   
V15 &   9:34:0.46  & 55:14:31.31 & 125.0    &~~~~~~3715.055&  CC   &23.65  &0.54  & 22.68   & 0.41 & 0.76\\
\hline
\end{tabular}
\end{center}
\normalsize
$^{a}$ Variable stars were given increasing numbers with increasing the distance from the
galaxy center which was set at $\alpha = 9^h 34^m 02^s.10$ and $\delta = 55^{\circ} 14^{\prime} 26^{\prime \prime}.81$ (J2000.0).\\
$^{b}$ Coordinates have been measured on the HST image j9ej01010\_drz.fits.\\
$^{c}$ CC = Classical Cepheid; LPV = long period variable.\\ 
$^{d}$ A$_V$ and A$_I$ indicate the amplitude of the light variation in the $V$ and $I$ bands, respectively.\\
\end{table*}
\linespread{1}

\begin{table*} \caption{Identification and properties of candidate variable 
stars in IZw18}
\begin{center}
\begin{tabular}{lcccccc}
\hline
\hline
          &   ${\rm \alpha }$ & ${\rm \delta}$  &  &  \\
            ${\rm Name^a}$& ${\rm (2000.0)}$& ${\rm (2000.0)}$ &${\langle V\rangle}$ & $\Delta F606W$ &
	    ${\langle I\rangle}$ & $\Delta F814W$\\
            \hline
 V2  &    9:34:01.76  &  55:14:24.75  &    28.17$\pm$0.10  &  0.90 &  28.16$\pm$0.16   &  0.86 \\ 
 V3  &    9:34:02.19  &  55:14:24.03  &    26.43$\pm$0.04  &  0.71 &  26.35$\pm$0.10   &  0.64 \\
 V5  &    9:34:02.31  &  55:14:24.25  &    26.15$\pm$0.04  &  0.65 &  25.75$\pm$0.16   &  0.70 \\
 V8  &    9:34:01.35  &  55:14:22.08  &    26.50$\pm$0.03  &  0.27 &  24.85$\pm$0.03   &  0.30 \\
 V9  &    9:34:02.65  &  55:14:21.37  &    26.79$\pm$0.11  &  0.90 &  25.83$\pm$0.16   &  0.81 \\
 V10 &    9:34:02.59  &  55:14:20.78  &    26.39$\pm$0.03  &  0.50 &  26.49$\pm$0.07   &  0.51 \\
 V11 &    9:34:02.67  &  55:14:35.87  &    28.38$\pm$0.06  &  1.46 &  27.80$\pm$0.11   &  0.95 \\ 
 V12 &    9:34:03.24  &  55:14:29.00  &    28.79$\pm$0.08  &  0.62 &  28.31$\pm$0.17   &  0.35 \\ 
 V13 &    9:34:03.06  &  55:14:20.61  &    27.73$\pm$0.04  &  0.66 &  26.59$\pm$0.07   &  0.71 \\
 V14 &    9:34:03.08  &  55:14:20.13  &    26.93$\pm$0.03  &  0.25 &  25.07$\pm$0.05   &  0.66 \\
 V16 &    9:34:02.74  &  55:14:42.78  &    27.17$\pm$0.03  &  0.87 &  25.18$\pm$0.05   &  1.24 \\
 V17 &    9:34:04.05  &  55:14:32.45  &    28.13$\pm$0.05  &  0.81 &  26.77$\pm$0.07   &  0.81 \\ 
 V18 &    9:33:59.67  &  55:14:21.82  &    27.41$\pm$0.06  &  1.11 &  26.77$\pm$0.13   &  1.02 \\
 V19 &    9:34:00.53  &  55:14:43.74  &    27.94$\pm$0.04  &  0.98 &  28.22$\pm$0.10   &  0.88 \\ 
 V20 &    9:34:00.19  &  55:14:42.60  &    27.36$\pm$0.03  &  0.69 &  26.89$\pm$0.05   &  0.67 \\
 V21 &    9:34:00.38  &  55:14:44.26  &    27.60$\pm$0.03  &  1.02 &  27.70$\pm$0.11   &  1.45 \\
 V22 &    9:34:00.14  &  55:14:43.05  &    26.21$\pm$0.03  &  0.34 &  26.13$\pm$0.07   &  0.66 \\
 V23 &    9:33:59.78  &  55:14:43.99  &    24.98$\pm$0.03  &  0.40 &  23.62$\pm$0.04   &  0.20 \\
 V24 &    9:34:02.25  &  55:14:52.83  &    27.50$\pm$0.04  &  0.35 &  27.08$\pm$0.13   &  0.35 \\ 
 V25 &    9:33:59.75  &  55:14:44.04  &    24.16$\pm$0.05  &  0.74 &  23.78$\pm$0.12   &  1.00 \\
 V26 &    9:33:59.68  &  55:14:43.64  &    28.32$\pm$0.06  &  1.12 &  27.81$\pm$0.12   &  0.93 \\ 
 V27 &    9:33:59.49  &  55:14:44.08  &    27.54$\pm$0.05  &  0.71 &  26.43$\pm$0.05   &  0.71 \\
 V28 &    9:33:58.60  &  55:14:32.22  &    27.56$\pm$0.06  &  0.90 &  26.70$\pm$0.12   &  1.15 \\
 V29 &    9:33:58.77  &  55:14:38.00  &    27.42$\pm$0.04  &  0.80 &  26.25$\pm$0.06   &  1.25 \\
 V30 &    9:33:58.51  &  55:14:38.07  &    27.60$\pm$0.05  &  0.83 &  26.68$\pm$0.11   &  0.75 \\
 V31 &    9:33:59.22  &  55:14:48.50  &    26.69$\pm$0.05  &  0.30 &  25.45$\pm$0.05   &  0.44 \\
 V32 &    9:33:59.46  &  55:14:51.53  &    28.11$\pm$0.06  &  1.12 &  26.78$\pm$0.09   &  1.26 \\
 V33 &    9:33:58.65  &  55:14:42.92  &    27.54$\pm$0.04  &  0.62 &  26.21$\pm$0.05   &  1.09 \\
 V34 &    9:33:58.87  &  55:14:49.45  &    27.84$\pm$0.05  &  0.70 &  26.97$\pm$0.10   &  1.69 \\
&&&&\\
\hline
\end{tabular}
\end{center}
$^{a}$ Variables were given increasing numbers with increasing the distance from the
galaxy center.\\
\end{table*}
\linespread{1}

\begin{table}
\begin{center}
\caption{$V,I$ photometry of the confirmed variable stars in IZw18}
\begin{tabular}{cccc}
\hline
\hline
\multicolumn{4}{c}{\rm Star V6 - Classical Cepheid - P=8.71 days}\\
\hline
{\rm HJD} & ${V}$  & {\rm HJD } & ${I}$\\ 
{\rm ($-$2450000)} &   & {\rm ($-$2450000) } &  \\
\hline
  2786.129397  &  27.33$\pm$0.08   &    2788.995991  &   26.29$\pm$0.08  \\  
  2790.301071  &  26.95$\pm$0.06   &    2790.995864  &   26.53$\pm$0.07  \\  
  2791.995813  &  27.44$\pm$0.06   &    2795.928907  &   26.05$\pm$0.09  \\  
  2793.795690  &  27.51$\pm$0.10   &    3646.200173  &   26.63$\pm$0.06  \\  
  2796.879372  &  26.64$\pm$0.04   &    3679.282332  &   26.64$\pm$0.07  \\  
  3646.074922  &  27.37$\pm$0.07   &    3683.411429  &   26.83$\pm$0.06  \\  
  3673.354491  &  27.53$\pm$0.07   &    3685.276123  &   26.29$\pm$0.12  \\  
  3679.215186  &  27.32$\pm$0.06   &    3688.189299  &   26.48$\pm$0.07  \\  
  3683.344260  &  27.26$\pm$0.06   &    3692.653076  &   26.33$\pm$0.11  \\  
  3685.208977  &  26.84$\pm$0.04   &    3708.381510  &   26.85$\pm$0.06  \\  
  3688.113912  &  27.32$\pm$0.10   &    3714.649539  &   26.58$\pm$0.09  \\ 
  3692.577572  &  26.98$\pm$0.05   &    3725.456858  &   26.63$\pm$0.08  \\ 
  3708.326761  &  27.55$\pm$0.10   &    3730.318190  &   26.22$\pm$0.06  \\ 
  3714.535979  &  27.30$\pm$0.06   &    3737.177722  &   26.28$\pm$0.06  \\ 
  3725.390374  &  27.59$\pm$0.09   &    3742.103508  &   26.84$\pm$0.06  \\ 
  3730.251196  &  26.96$\pm$0.04   &     --          &     --        \\
  3737.110567  &  26.63$\pm$0.03   &     --          &     --        \\
  3742.035879  &  27.30$\pm$0.05   &     --          &     --        \\
\hline

\end{tabular}

\end{center}
\medskip

A portion of Table\,4 is shown here for guidance regarding its form
and content. The entire catalog is available in the electronic edition 
of the journal.
\end{table}
\normalsize

\clearpage

\end{document}